\documentclass[prl,superscriptaddress,twocolumn,showpacs,epsf]{revtex4}
\usepackage{epsfig}
\usepackage{color}

\usepackage{graphicx}
\usepackage{latexsym}
\usepackage{amsmath}
\usepackage{amssymb}
\usepackage{amsfonts}

\begin{document}

\title{Tunable anisotropic nonlinearity in superconductors with asymmetric antidot array}

\author{A. Yu. Aladyshkin}
\affiliation{INPAC -- Institute for Nanoscale Physics and
Chemistry, Nanoscale Superconductivity and Magnetism Group,
K.U.Leuven, Celestijnenlaan 200D, B--3001 Leuven, Belgium}
\affiliation{Institute for Physics of Microstructures RAS, 603950
Nizhny Novgorod, GSP-105, Russia}
\author{J. Van de Vondel}
\affiliation{INPAC -- Institute for Nanoscale Physics and
Chemistry, Nanoscale Superconductivity and Magnetism Group,
K.U.Leuven, Celestijnenlaan 200D, B--3001 Leuven, Belgium}
\author{C. C. de Souza Silva}
\affiliation{Departamento de Fisica, Universidade Federal de
Pernambuco, Cidade Universitaria, 50670-901 Recife-PE, Brazil}
\author{V. V. Moshchalkov}
\affiliation{INPAC -- Institute for Nanoscale Physics and
Chemistry, Nanoscale Superconductivity and Magnetism Group,
K.U.Leuven, Celestijnenlaan 200D, B--3001 Leuven, Belgium}

\date{\today}
\begin{abstract}
The influence of the spatial asymmetry of the pinning potential on
the spectral composition of the voltage, induced in perforated
superconducting Al bridges by the injection of a sinusoidal bias
current, was investigated. The loss of the mirror
symmetry of the pinning potential leads to the appearance of even
Fourier components in the induced voltage in the vicinity of the
superconducting phase transition line on the $H-T$ diagram ($H$ is
the external magnetic field, $T$ is temperature).
Artificially-introduced asymmetry for vortex motion makes it
possible to create low-resistive materials, in which nonlinearity
depends on the direction of injected electrical currents.
\end{abstract}

\maketitle
\newpage

The controlled manipulation of magnetic flux quanta (or vortices)
in superconducting films under the action of non-equilibrium
fluctuations or external excitations with zero average (vortex
ratchet) has recently attracted a lot of interests (see, e.g.,
Refs.~\cite{Lee-99,Wambaugh-99,Olson-01,Zhu-03,Carneiro-05,Lu-2007,Villegas-03,Vondel-05,Togawa-05,Souza-06a,Vondel-07,Schildermans-07,Souza-07,Silhanek-07,Gillijns-07}
and references therein). A preferable vortex motion in a certain
('easy') direction can be experimentally realized in various
one-dimensional (1D) and two-dimensional (2D) systems of reduced
symmetry: in films modulated by triangular
dots~\cite{Villegas-03}, in perforated films with regular arrays
of asymmetric
antidots~\cite{Vondel-05,Togawa-05,Souza-06a,Vondel-07}, in
mesoscopic singly-connected
superconductors~\cite{Schildermans-07}, and in superconductors
placed in a spatially-modulated magnetic
field~\cite{Souza-07,Silhanek-07,Gillijns-07}. All these systems
can be potentially used for designing flux pumps for removal of
unwanted trapped flux in superconducting devices \cite{Lee-99},
and dc rectifiers (diodes).

In this Letter we are aiming at the generalization of the concept
of unidirectional vortex dynamics in a ratchet potential for the
case of {\emph{ higher harmonics generation}}. Indeed, passing an
alternating current of certain frequency $f$ through any medium
should generally induce voltage oscillations at multiple
frequencies $nf$ ($n$ is integer). However, the appearance of even
harmonics (regardless on the frequency range) is known to be
forbidden for a material with spatial inversion or mirror plane
symmetry \cite{LandauLifshitz}. One can expect that an
artificially-introduced asymmetry for vortex motion, occurring in
all ratchet systems, will lead to a nontrivial nonlinear response.
In the presence of a ratchet potential the static $I-V$ dependence
of the perforated superconducting bridge will lose symmetry,
$V(I)\neq-V(-I)$. Since the voltage drop in a sample should be
zero at $I=0$, the expansion of $V$ vs. $I$ in power series has
the following form: $V=\alpha_1 I+\alpha_2 I^2+\alpha_3 I^3+... $,
where $\alpha_i$ are constants depending on the external magnetic
field $H$ and temperature $T$. Substituting $I(t)=I_0\sin(2\pi
ft)$, one can see that the time-dependent voltage $V(t)$
oscillates at multiple frequencies, including zero frequency.
Provided $V(I)=-V(-I)$, both the diode effect and the even Fourier
components should disappear. It is worth noting that breaking the
sample symmetry in 2D superconducting films can affect only a
certain direction, while keeping the symmetry in the perpendicular
direction unaltered. It gives us a possibility to fabricate
{\emph{ an artificial material with anisotropic nonlinearity,
depending on the orientation of the injected electrical
currents}}.

\begin{figure}[b]
\includegraphics[width=5.5cm]{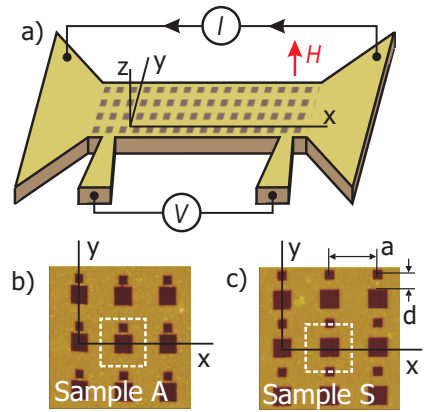}
\caption{(Color online) (a) A schematic view of the perforated
superconducting Al bridge; (b) and (c) Atomic force microscopy
images of samples A and S. White dashed lines depict the unit
cells ($3\times 3~\mu$m$^2$).} \label{Fig-System}
\end{figure}

\begin{figure}[t]
\includegraphics[width=7.5cm]{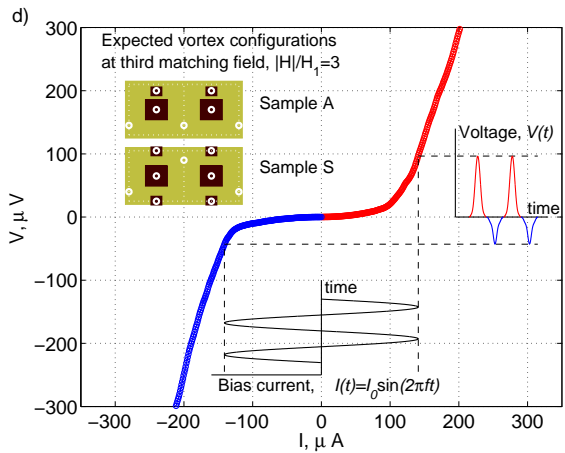}
\caption{(Color online) An example of the non-symmetrical static
$I-V$ dependence, measured at $T=1.285$~K and $H=-0.68$~mT (close
to the third matching field) for the sample A. The insets show the
transformation of the sinusoidal bias current into a complex
signal with nonzero average (the diode effect).} \label{Fig-IV}
\end{figure}

Two perforated Al microbridges of the width 600 $\mu$m and
thickness $d=50$ nm, which have the critical temperatures $T_{c}$
close to 1.3~K, were prepared by dc-sputtering on a Si/SiO$_2$
substrate. The pinning landscape, prepared with electron beam
lithography, has the period $a=3000~$nm and it can be represented
as a superposition of two interpenetrated square sublattices of
big and small microholes ("antidots") of
\mbox{$1200\times1200$~nm$^2$} and \mbox{$600\times600$~nm$^2$} in
size (Fig.~\ref{Fig-System}). Depending on the separation $s$
between centers of these antidots the resulting complex array can
be symmetrical ($s=1500$~nm) or asymmetrical ($s=1100$~nm) with
respect to the $y-$axis. For brevity we introduced notations
"sample S" and "sample A" (either symmetrical or asymmetrical
pinning potential).

\begin{figure*}[bth!]
\includegraphics[height=6.5cm]{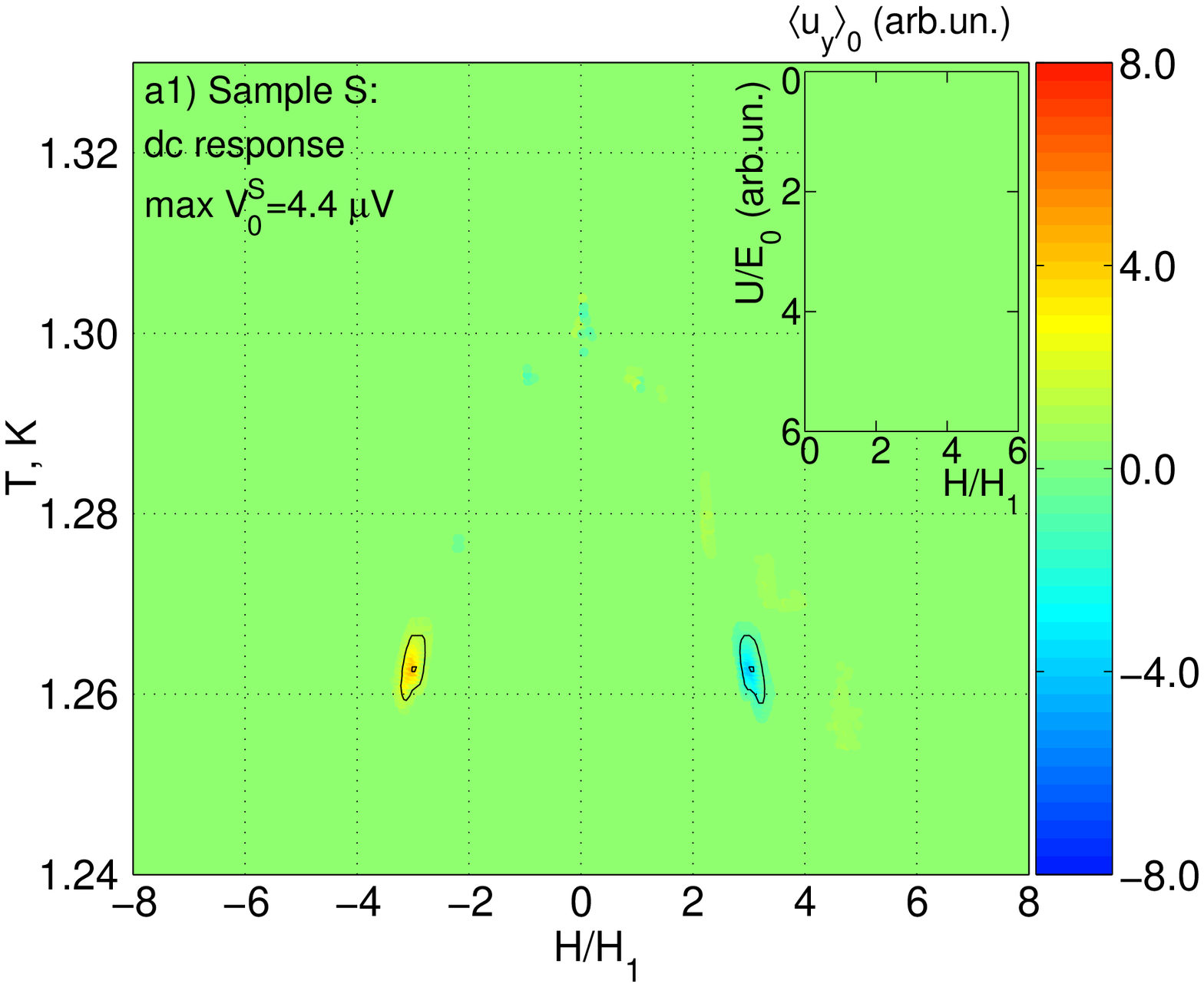}
\includegraphics[height=6.5cm]{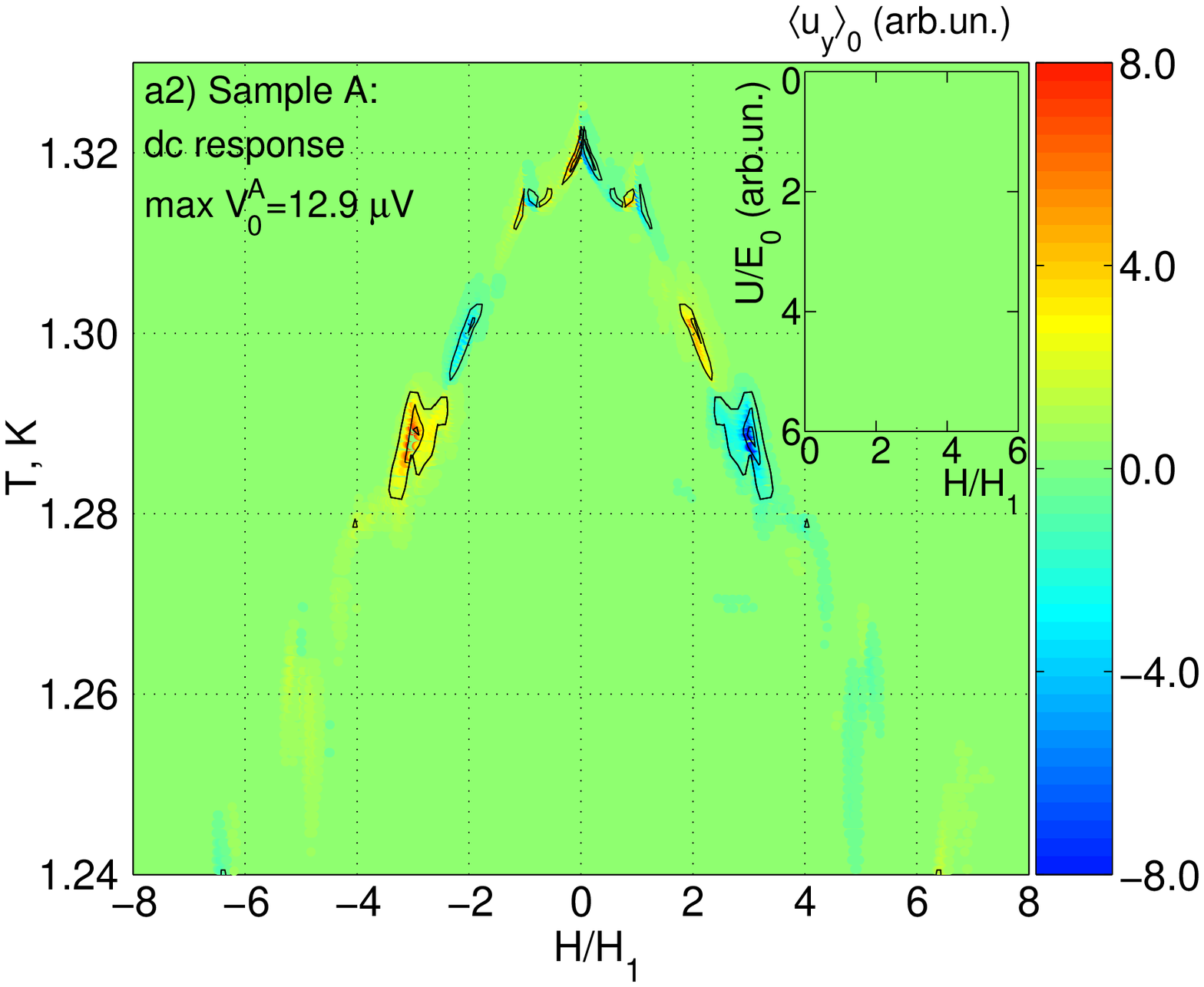}
\includegraphics[height=6.5cm]{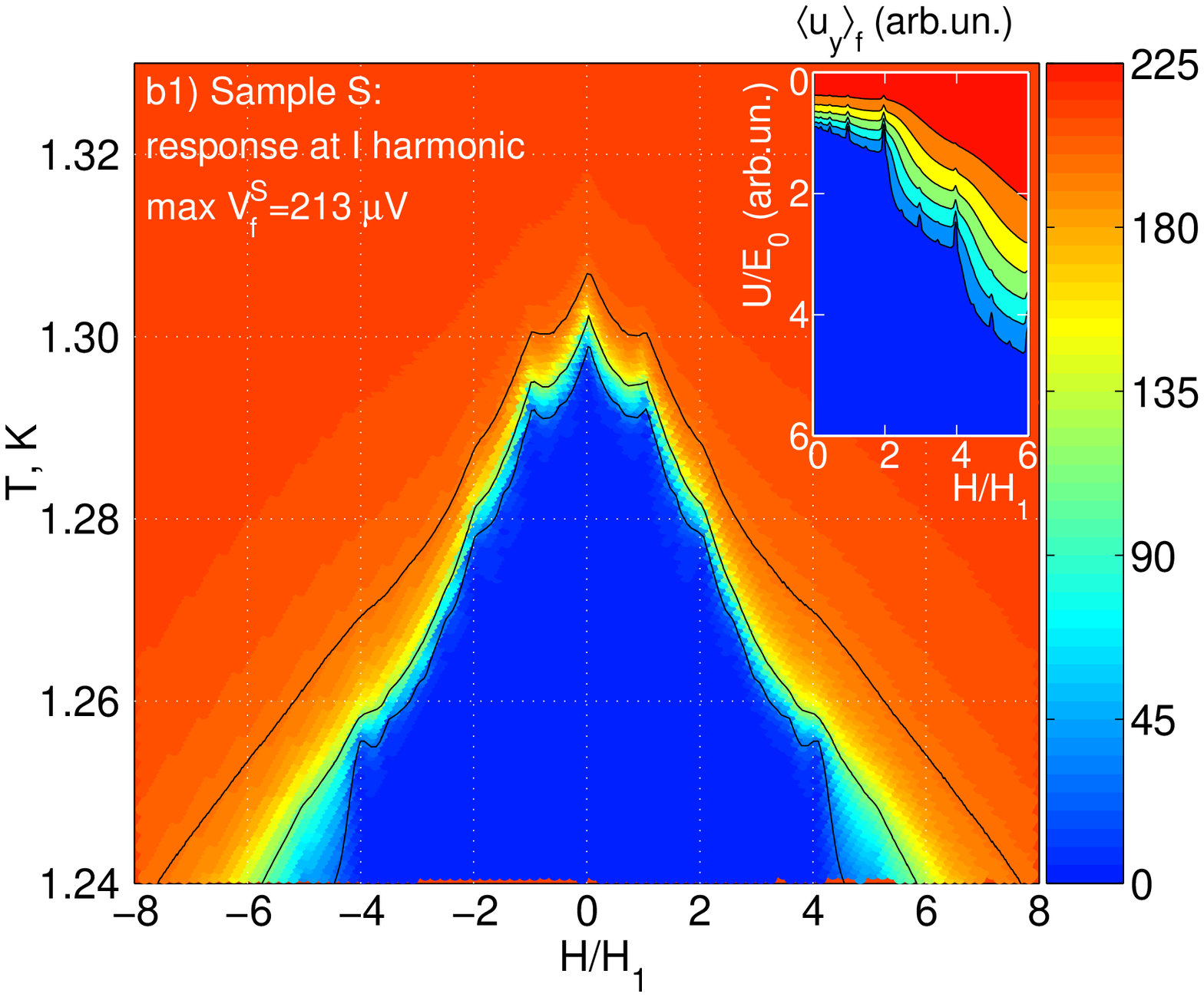}
\includegraphics[height=6.5cm]{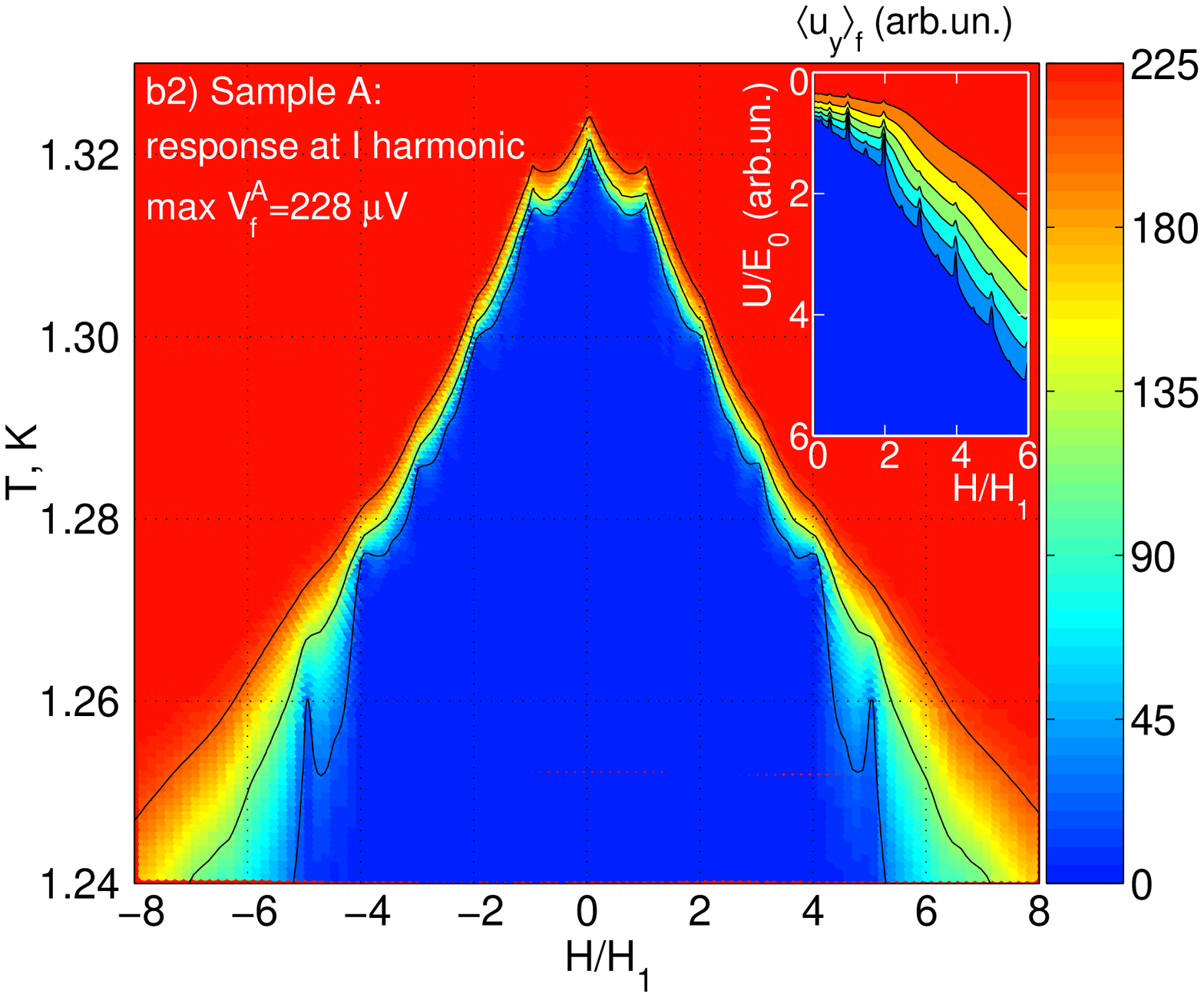}
\caption{(Color online) The dependencies of the voltage $V_{nf}$,
measured at zero and fundamental on the external magnetic field
$H$ and temperature $T$: (a1) $V^{S}_0$, (a2) $V^{A}_0$, (b1)
$V^{S}_f$, (b2) $V^{A}_f$. Solid black lines correspond to the
conditions $|V_{nf}(H,T)|=\alpha\,\max V_{nf}$, $\alpha=0.1, 0.5,
0.95$. All insets show the dependencies of the Fourier components
of the mean instant velocity $\langle u_y^S \rangle_{nf}$ and
$\langle u_y^A \rangle_{nf}$ on $H$ and the effective inverse
temperature $U/E_0$, calculated with the assumption of the
logarithmic intervortex interaction $V(r)=-E_0\,\ln(r/\lambda)$,
where $U$ is the pinning energy. Note that the periodic changes in
$\langle u^{A}_y \rangle_{0}$, $\langle u^{A}_y \rangle_{f}$ and
$\langle u^{S}_y \rangle_{f}$ at sweeping $H$ resemble the
variations of the Fourier components of the voltage. Color scales
for all plots and insets are chosen in the same way.}
\label{Fig-Experiment1}
\end{figure*}

\begin{figure*}[bth!]
\includegraphics[height=6.5cm]{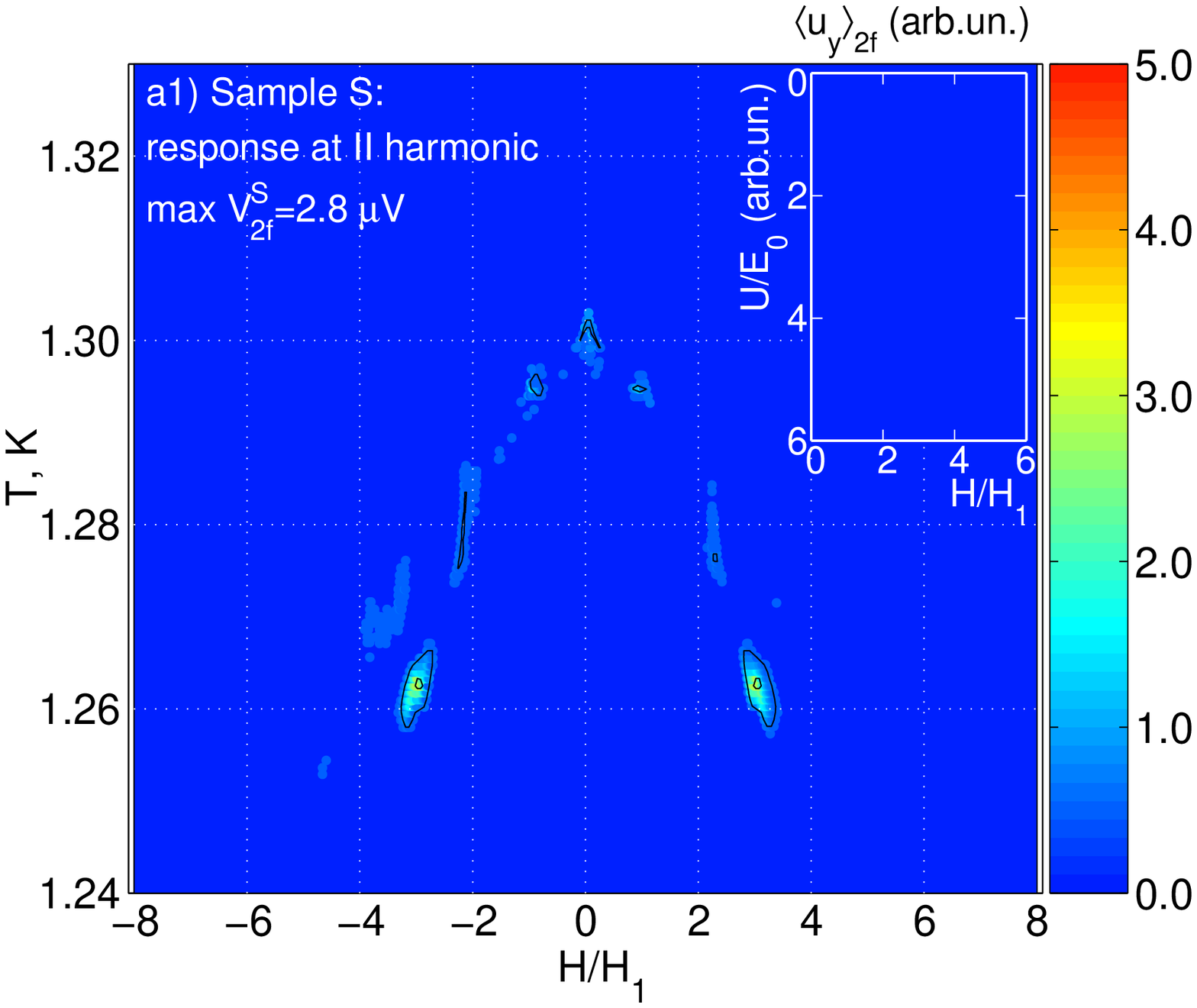}
\includegraphics[height=6.5cm]{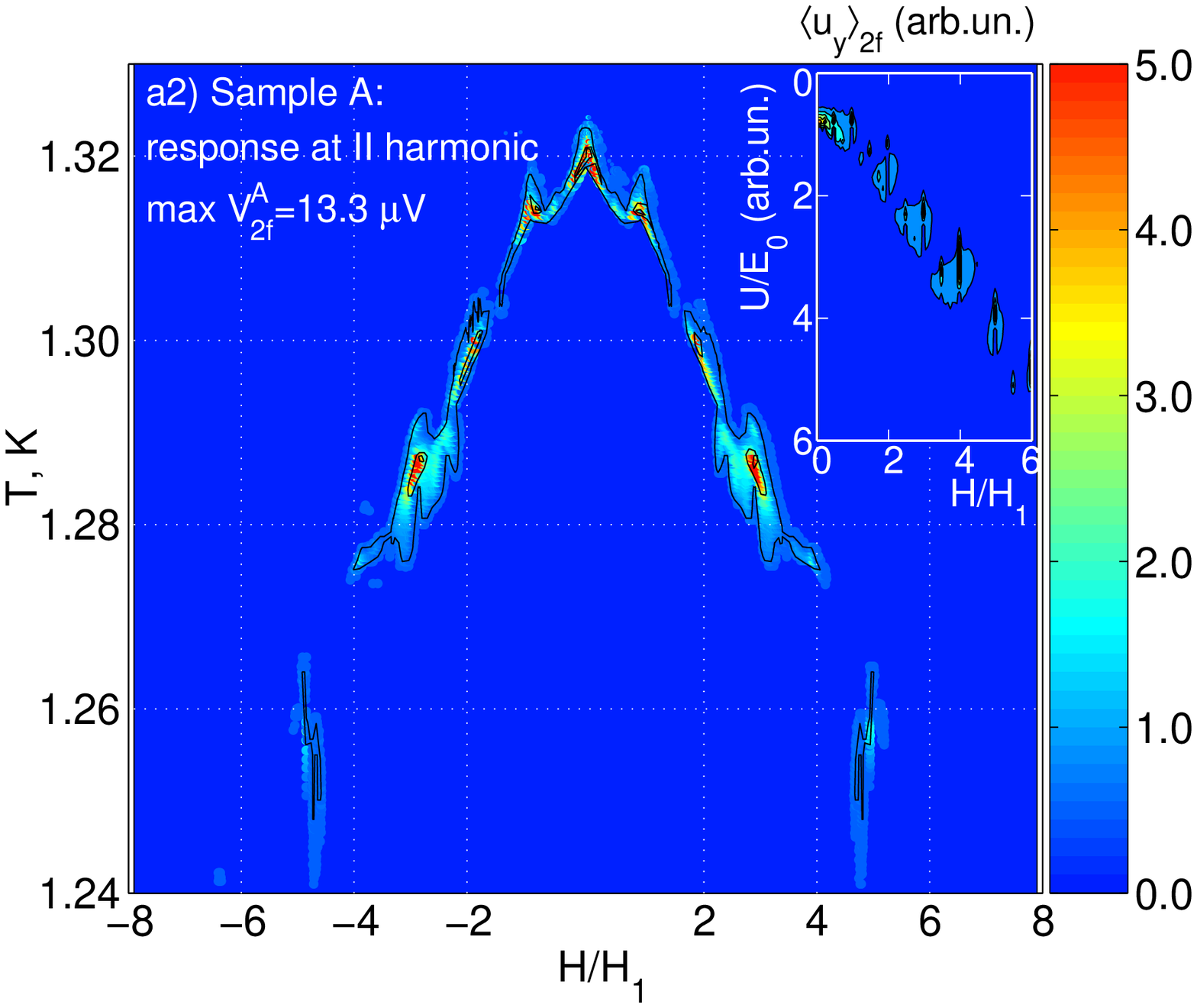}
\includegraphics[height=6.5cm]{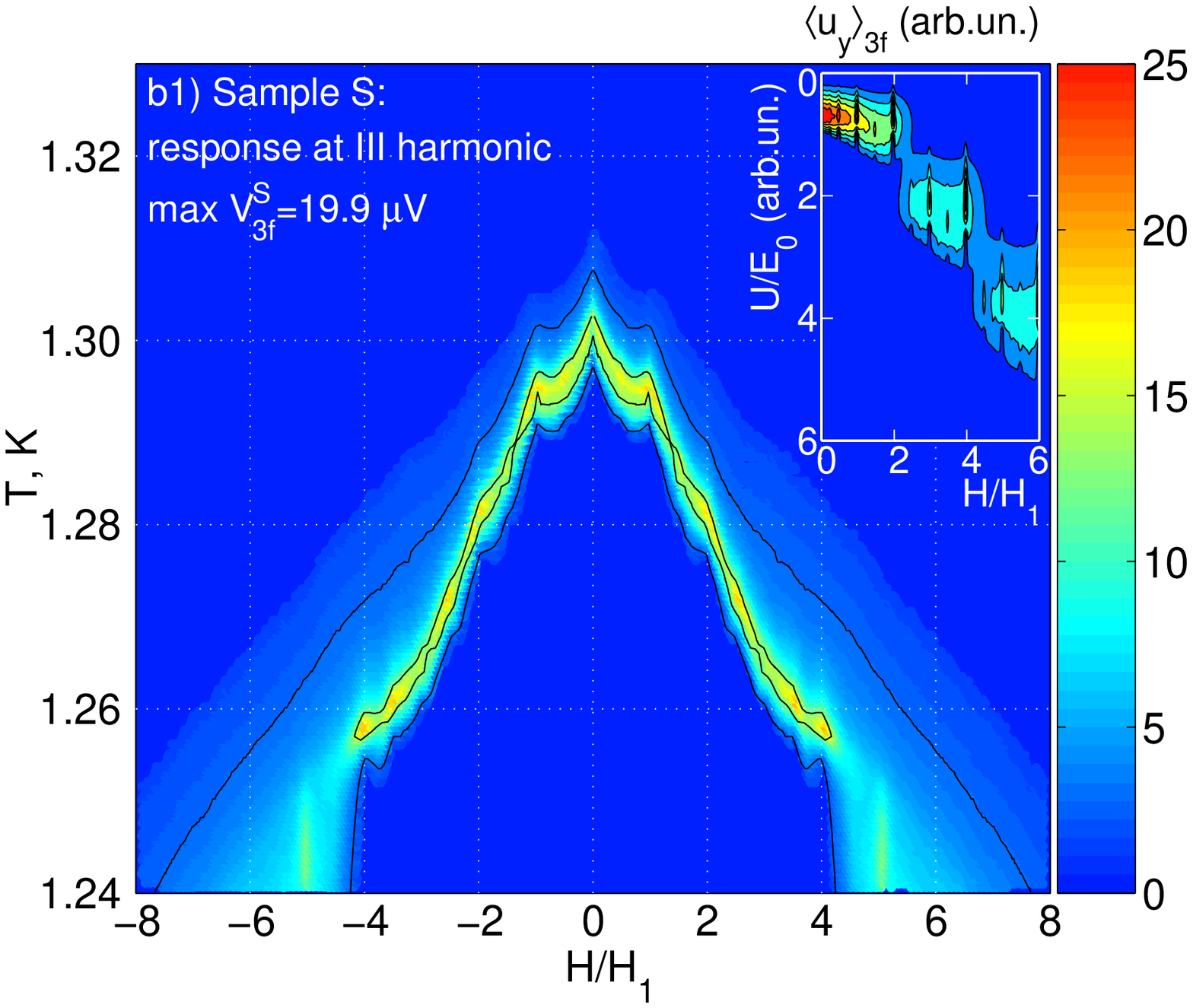}
\includegraphics[height=6.5cm]{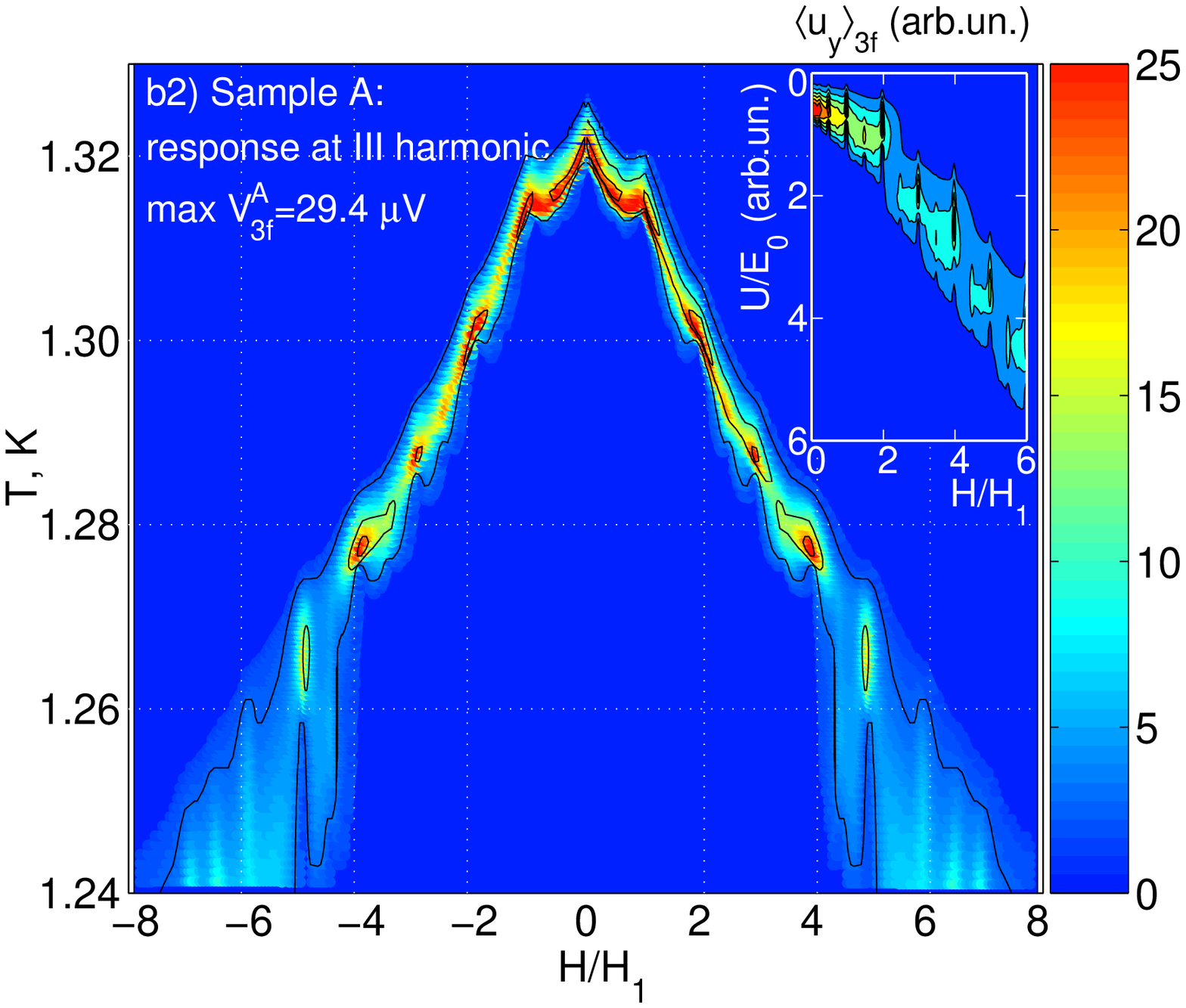}
\caption{(Color online) The dependencies of the voltage $V_{nf}$,
measured at double and triple frequencies, on the external
magnetic field $H$ and temperature $T$: (a1) $V^{S}_{2f}$, (a2)
$V^{A}_{2f}$, (b1) $V^{S}_{3f}$, (b2) $V^{A}_{3f}$. All notations
are the same as in Fig. \ref{Fig-Experiment1}.}
\label{Fig-Experiment2}
\end{figure*}

Our ac four-probe measurements were carried out as follows: a
sinusoidal driving current \mbox{$I=I_0\sin(2\pi ft)$}, applied to
such a bridge (the amplitude $I_0=141~\mu$A and the frequency
$f=1.11$~kHz), generates a voltage drop $V(t)$. The typical $I-V$
dependence is shown in Fig.~\ref{Fig-IV}. This time-dependent
voltage can be represented as $V(t)=\sum\limits_{n=0}^{\infty}
V_{nf}\,\sin(2\pi n f t+\varphi_n)$, where $V_{nf}$ and
$\varphi_n$ are the amplitude and the phase of the $n-$th Fourier
component. The amplitudes of the first, second and third Fourier
harmonics $V_f$, $V_{2f}$ and $V_{3f}$ as well as the dc voltage
$V_0$ as functions of the transverse magnetic field $H$ and
temperature $T$ were measured by the lock-in amplifier Signal
Recovery 7225 and the nanovoltmeter Hewlett Packard 34420A. All
experimental results are summarized in Figs.~\ref{Fig-Experiment1}
and \ref{Fig-Experiment2}.

We begin with the discussion of the $H-T$ diagrams of the linear
response $V_f^S$ and $V_f^A$, measured for samples S and A at the
fundamental frequency [panels (b1) and (b2) in
Fig.~\ref{Fig-Experiment1}]. The occurrence of resistivity minima
at integer occupation numbers \mbox{$m=H/H_1$} in both samples
indicates the formation of very stable vortex configurations. Here
$H_1=\Phi_0/a^2\simeq 0.22$~mT is the first matching field, at
which the sample is threaded by one magnetic flux quantum,
$\Phi_0=2.07\cdot10^{-15}$ Wb, per unit cell. The broader
resistive transition of the sample S in comparison with the sample
A (at the same $T/T_c$ ratio) can be attributed to the enhanced
surface superconductivity, since the antidots in the sample S are
distributed more uniformly across the sample area.

As expected, the loss of mirror symmetry of the pinning landscape
along the $y-$axis significantly increases the dc rectification
for sample A [compare the panels (a1) and (a2) in
Fig.~\ref{Fig-Experiment1}]. The voltage pattern in the $H-T$
diagram demonstrates the series of sign inversions which were
recently explained in Ref.~\cite{Souza-06a} by taking into account
the intervortex interaction for a dense vortex lattice in the
presence of a ratchet potential.  Under the same conditions the
diode effect for sample S is practically absent, except in the
vicinity of the third matching field ($|H|/H_1\simeq 3$). The
origin of such 'resonant' rectification for the nominally
symmetrical sample is currently unclear.

The $H-T$ diagrams of the response at the second Fourier harmonic,
$V^{S}_{2f}$ and $V^{A}_{2f}$ seem to be similar to that for the
dc response in many respects [compare the panels (a1) and (a2) in
Figs.~\ref{Fig-Experiment1} and \ref{Fig-Experiment2}]. Due to a
larger "signal-to-noise" ratio for measurement of the second
Fourier harmonic in comparison to dc measurements, this modulation
measurements appear to be more effective for a probing the
symmetry properties of samples. Finally, we showed the results of
the measurements at the triple frequency, $V^{S}_{3f}(H,T)$ and
$V^{A}_{3f}(H,T)$ [panels (b1) and (b2) in
Fig.~\ref{Fig-Experiment2}]. We emphasize that the field-induced
oscillations of the critical temperature, related to the formation
of stable vortex structures, become more pronounced as compared
with the $T_c$ oscillations observed in the linear response. Thus,
the experimental treatment of the nonlinear response is a rather
sensitive and powerful technique for studying the commensurability
effects in nanostructured superconductors.

In order to illustrate correlations between the symmetry of the
pinning potential and the spectral composition of the induced
voltage, we simulated the low-frequency dynamics of a vortex chain
in an asymmetrical 1D periodic potential~\cite{Souza-06a} under
the sinusoidal current excitation and estimated the instant mean
vortex velocity $\langle u_y(t) \rangle=\sum\limits_{i=1}^{N}
\dot{y}_i/N$ ($N$ is the number of vortices). We considered both
long-range vortex-vortex interaction, $V(r)=-E_0\,\ln(r/\lambda)$,
and short-range one, $V(r)=E_0\,K_0(r/\lambda)$ ($K_0$ is the
modified Bessel function), used recently for describing ratchet
phenomena in superconducting systems with the period comparable
with the penetration length $\lambda$ \cite{Gillijns-07}. The
Fourier components of the instant mean vortex velocity, $\langle
u_y \rangle_{nf}=\left|\int_0^{1/f}\!\langle u_y(t)
\rangle\,e^{-2\pi inft}\,dt\right|$, as well as its dc value,
$\langle u_y \rangle_{0}=\int_0^{1/f}\!\langle u_y(t)
\rangle\,dt$, are expected to determine the main contribution to
the spectrum of the voltage: $V_{0}\propto H\langle u_y
\rangle_{0}$ and $V_{nf}\propto H\langle u_y \rangle_{nf}$. It was
found that independently on the type of the intervortex
interaction, this simple model does explain how an ac excitation
of interacting vortices in the ratchet potential leads to a
generation of even Fourier harmonics observed experimentally. The
results of our simulation are plotted in the $H/H_1-U/E_0$ plane
(see the insets in Figs. \ref{Fig-Experiment1}
\ref{Fig-Experiment2}), where $U$ is the characteristic pinning
strength. Since $U_p$ is known to decrease at increasing
temperature and vanish at $T=T_{c}$ \cite{Blatter}, we can
consider a wide range of the pinning strengths (and temperatures),
starting from $U_p=0$. It should be noted that our 1D model cannot
describe the rectification at $|H|\simeq 3H_1$ for the sample S.
This indicates out that the dynamics of 2D vortex lattice in the
2D pinning potential is crucial for the explanation of all
observed peculiarities in nonlinear response. However this issue
goes beyond the subject of this Letter, and the detailed treatment
in the framework of a full 2D model similar to that used in Refs.
\cite{Lu-2007,Gillijns-07} will be considered separately.

Summarizing, we presented a comparative study of the linear and
nonlinear properties of the perforated superconducting films. We
demonstrated that the nonlinear properties of the nanostructured
superconductors can be tuned by making the pinning landscape of a
required symmetry, and thus we proposed a tunable vortex frequency
converter. Interestingly, such a symmetry-breaking induced
harmonics generation could be a basis for a designing the
materials with an anisotropic nonlinearity.

This work was supported by the K.U. Leuven Research Fund
GOA/2004/02 program, NES -- ESF program, the Belgian IAP, the Fund
for Scientific Research -- Flanders (F.W.O.--Vlaanderen), the
Brazilian research agency CNPq, the Russian Foundation for Basic
Research, RAS under the program "Quantum physics of condensed
matter" and the Presidential grant MK-4880.2008.2.

\end{document}